\documentclass[aps,pra,amsmath,amssymb,11pt,final,tightenlines,twoside,onecolumn,nofloats,nofootinbib,superscriptaddress,
showkeys,showkeywords]{revtex4-1}

\usepackage[T2A]{fontenc}
\usepackage[utf8x]{inputenc}
\usepackage[russian,english]{babel}
\usepackage{graphicx}

\begin{document}

\selectlanguage{english}

\noindent {\it ASTRONOMY REPORTS, 2026, Vol. , No. }
\bigskip\bigskip  \hrule\smallskip\hrule
\vspace{35mm}

\title{Modeling of the magnetic stellar wind braking of the
ssrAp 33 Lib (HD137949)}

\author{\bf\copyright~2025 A.G.~Nikiforov}
\email{anikiforov@inasan.ru}

\affiliation{Institute of Astronomy, Russian Academy of Sciences \\
119017 Moscow, 48 Pyatnitskaya St.}

\author{\bf V.D.~Bychkov}
\email{vbych@sao.ru}

\affiliation{Special Astrophysical Observatory, Russian Academy of Sciences \\
    369167
    Nizhny Arkhyz,
    Zelenchuksky District,
    Karachay-Cherkess Republic,
    Russia}

\author{\bf M.V.~Barkov}
\email{barkov@inasan.ru}

\affiliation{Institute of Astronomy, Russian Academy of Sciences \\
119017 Moscow, 48 Pyatnitskaya St.}

\date{\today}

\begin{abstract}
\vspace{3mm}
\received{12.06.24}
\revised{12.06.24}
\accepted{12.06.24} 
\vspace{3mm}
Using the ssrAp star 33 Lib (HD137949) as an example, we show that magnetic braking caused by the interaction of a strong magnetic field with a stellar wind can play a key role in slowing the rotation of ssrAp stars. Numerical modeling of stellar rotation spin-down in the \texttt{MESA} package, taking into account the evolution of magnetic fields and the stellar wind, shows that stars with rotation periods of up to 80 years and longer can form.
Moreover, braking by a magnetized wind makes it possible to estimate the mass-loss rate for stars of moderate mass ($1.25 M_\odot< M <2 M_\odot$), which is difficult to do by other methods. We introduce the dimensionless parameter $\Xi$, which reflects the spin-down time and the stellar lifetime. Thus, when $\Xi\gg1$, braking is important, whereas when $\Xi\ll1$, it is negligible.



\begin{description}
\item[Keywords]
stars, magnetic fields, rotation, stellar evolution
\end{description}
\end{abstract}


\maketitle

\section{Introduction}

Stellar magnetic fields play an important role in the physics and evolution of stars. 
Approximately 10\% of upper-main-sequence (MS) stars possess sufficiently strong magnetic fields \cite{2019MNRAS.483.2300S, 2017MNRAS.465.2432G} ($\ge 100$ G), 
reaching several thousand gauss in strength
\cite{2005A&A...430.1143B, 2021A&A...652A..31B}.
The observed longitudinal magnetic fields of these stars vary with rotational phase,
which is well described by the oblique rotator model; see \cite{stibbs1950study,1950ApJ...112..222S,1960ApJ...132..521B,1971PASP...83..571P,2000BaltA...9..253K}.
According to \cite{2021A&A...652A..31B}, 
75\% of stars exhibit harmonic magnetic phase curves and
25\% exhibit double-wave phase curves.
In most cases the magnetic fields remain stable over the entire observational interval
(about 75 years), although slight variability is suspected in only a few cases
\cite{2021A&AT...32..143B}. 
Strong magnetic fields stabilize the atmospheres of hot stars
\cite{1970ApJ...159..985D}, 
which allows the mechanism of selective diffusion
to operate efficiently in their atmospheres and to produce localized regions (``spots'') with enhanced abundances,
as a rule, of rare-earth elements and iron-peak elements
\cite{1981A&A...103..244M,2015MNRAS.454.3143A,2019MNRAS.482.4519A,2024MNRAS.52710376P,2024Galax..12...55R}. 
Another remarkable property of magnetic Ap stars is
the observed rapid photometric variability with periods from 3 to 15 minutes.
This is associated with pulsations in certain localized regions (spots). Globally,
stars of such mass, size, etc. cannot pulsate in this way \cite{Kurtz1982}. The minimum period
of global pulsations in such stars would be at least 1.5 hours. Pulsations
in local regions
are associated with the spotted surface structure and should not affect the loss
of angular momentum in any way. This phenomenon may provide yet another additional
argument in favor of solid-body rotation in these stars.
The ``spots'' on the surfaces of Ap stars are characterized not only by overabundances of selected elements,
but also by a different surface temperature (more often lower), which leads to modulation
of the radiation with the rotation period.
Photometric variability provides an excellent opportunity to determine the rotation period of these stars
and many other parameters with high precision \cite{stibbs1950study, renson2001fifth, metlova2014photometric} 
.

The possibilities for determining the rotation periods of these stars have been greatly
expanded by the high-precision photometry of the Kepler, TESS, and other space missions \cite{balona2019rotational,david2019magnetic, sikora2019mobster, labadie2023photometric}.
Accurate knowledge of the periods greatly facilitates the construction of magnetic phase curves (MPCs) \cite{2023AstBu..78..141Y},
and thus increases the statistics of observed periods and MPC parameters. This will make it possible
to refine in the future the scenarios explaining the origin of global stellar magnetic fields.
However, one should point out a certain selectivity of this approach. It is connected with the fact
that the possibility of obtaining highly accurate photometric estimates from space missions such as
TESS, Kepler, Gaia, and others has appeared only relatively recently. As a consequence, as a rule,
it is difficult to search for and determine accurately periods longer than 50 days \cite{mathys2024long}.
At present, space missions allow estimates to be obtained only for
relatively short periods.

One of the most characteristic features of magnetic Ap stars is their
slow rotation. On average, Ap stars rotate 4--5 times more slowly
than normal stars of the same type \cite{2024AstBu..79..137G, 2021AstBu..76...91G, abt1995relation}. 
The possibility of rotation periods of
hundreds and even thousands of years was pointed out in \cite{mathys2017ap}. 
This statement is confirmed by observational results reported in \cite{hubrig2018magnetic, giarrusso2022twenty, bychkov2016periods, metlova2014photometric}
and a number of other works. This made it possible to identify a group of super-slowly rotating Ap stars---
ssrAp stars (super-slowly rotating Ap stars).
The main directions proposed to explain this observed effect of slow
rotation are as follows:

1. Stars arrive on the main sequence
having already lost a significant fraction of their angular momentum \cite{mouschovias1979angular}. \

2. Magnetic stars lose angular momentum during their main-sequence lifetime
due to the interaction of the magnetic field
of a rotating magnetic star with the interstellar medium; \cite{hartoog1977rotation,havnes1971magnetic,1988mast.conf..241F}
this interaction was considered through the operation of the Illarionov--Sunyaev propeller mechanism
\cite{illarionov1975number}. \ 

3. The possibility of efficient braking by a stellar wind was also considered
for outflowing matter from the magnetic poles of relatively hot Ap/Bp stars. The outflow of ionized
matter was controlled by the global dipolar magnetic field of the star. \

The observed properties of ssrAp stars, such as quasi-periodic brightness oscillations, spotted element distributions, and the synchronization of magnetic and rotation axes, are directly related to the dynamics of their magnetic fields and to rotational spin-down. However, the nature of the braking remains insufficiently studied: traditional models based on energy dissipation through a stellar wind are poorly consistent with the extremely low mass-loss rates of these objects. Alternative hypotheses suggest that large-scale magnetic fields play a dual role---not only suppressing meridional mixing and convection, but also redistributing angular momentum throughout the star through interaction with its ionized layers.

This work focuses on analytical estimates based on observational data and on modeling in order to study the connection between magnetic braking, wind, and the rotational dynamics of ssrAp stars. The results will make it possible to refine the braking parameters, improve the calibration of stellar ages, and explain their anomalous prevalence in old galactic populations.

In this work we used a solid-body model for stellar braking, because the Alfv\'en time for a star of this type is many orders of magnitude shorter than its braking time. A dipolar magnetic field was used. We also obtained an estimate of the mass-loss rate due to the stellar wind consistent with the estimate of the angular-momentum loss timescale.

The paper is organized as follows. Section 2 describes the analytical and numerical model of the magnetic braking mechanism. Section 3 presents the calculation results. Section 4 presents the conclusions and a discussion of the results.

\section{Magnetic Braking}

Magnetic braking of stars in the \texttt{MESA} code is described in \cite{Meynet2011}. The code takes stellar metallicity into account during main-sequence evolution; in our work solar metallicity was adopted ($\left[Fe/H\right]=0.02$).

The case of solid-body rotation differs radically from differential rotation, which is explained by the difference in the physical mechanisms responsible for the transport of angular momentum and chemical elements. In the solid-body rotation model, strong coupling between stellar layers and suppression of mixing are assumed, which allows stratification processes to develop in the radiative atmospheres of A stars. This contrasts with the differential-rotation scenario, in which chemical elements are homogenized by shear turbulence and meridional mixing.


The solid-body rotation model, even in the absence of magnetic braking, leads to a decrease in the angular momentum of the core and to its redistribution to the outer layers. Calculations performed in \cite{heger2005} demonstrate that within this model one can obtain neutron stars with initial rotation rates close to the upper limit of the observed values. At the same time, including magnetic braking in the calculation leads to a sharp additional decrease in the angular momentum of the core by orders of magnitude: for example, at a surface magnetic field strength at the magnetic equator of $B_{eq} = 100$ G, the value of $J$ decreases by more than an order of magnitude, and at 1 kG by 4 orders of magnitude.

\begin{figure}
    \centering
    \includegraphics[width=1.0\textwidth]{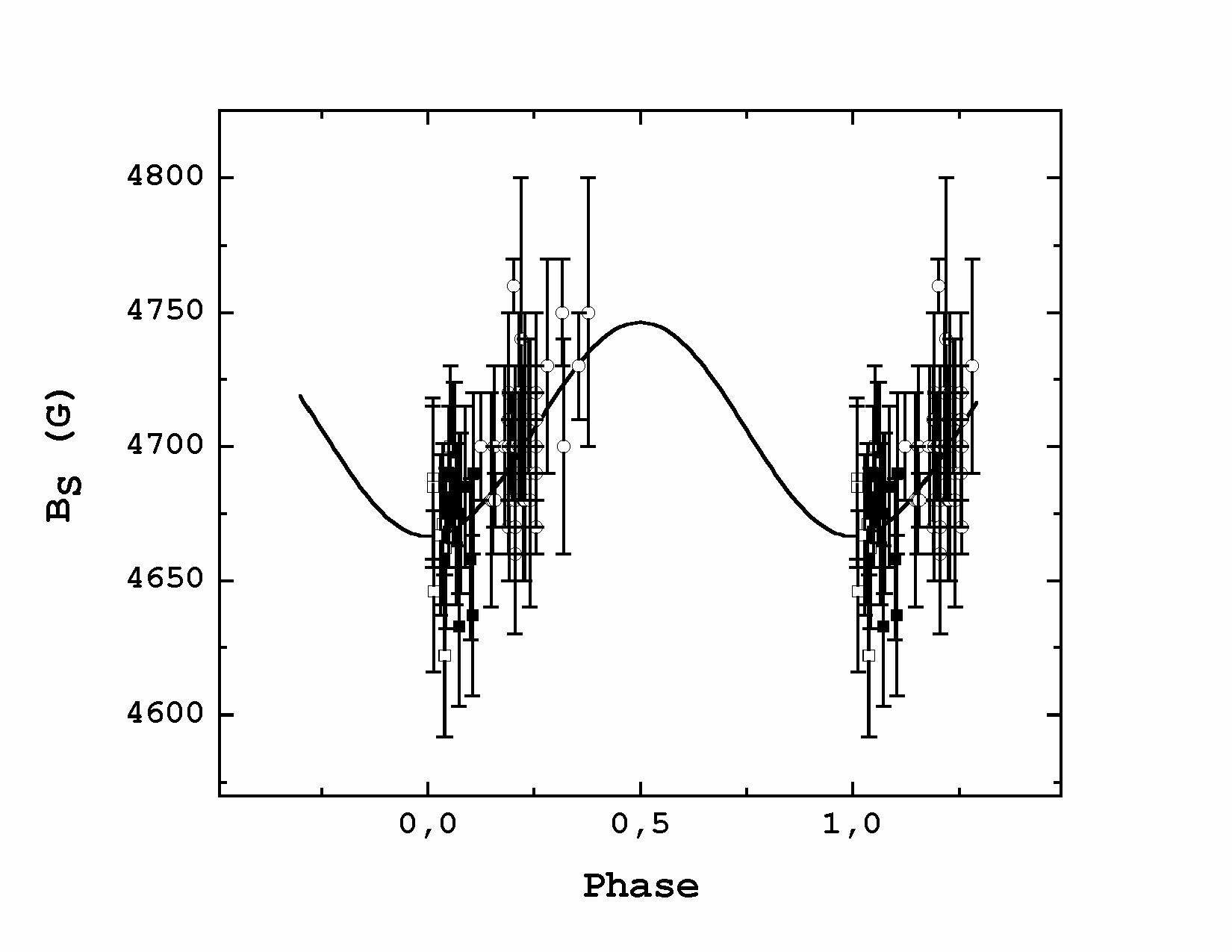}
    \caption{To construct the mean magnetic phase curve, estimates of the
    surface magnetic field from \cite{1997A&AS..123..353M,2017A&A...601A..14M,
    2022MNRAS.514.3485G} were used. }
    \label{fig:MFK}
\end{figure}

Let us consider the possible evolution of the rotation period using the typical ssrAp star 33 Lib (HD137949) as an example. The presence of a long rotation period (about 100 years) for this star had already been suggested in \cite{2014A&A...572A.113L}.
One of the probable rotation periods of this star was estimated to be 83.5 years (30478 days)
on the basis of a series consisting of 21 estimates of the longitudinal magnetic-field component $B_l$, obtained over
a time interval of 7528 days \cite{2015arXiv150606234B}.
Later, a more accurate estimate
of the period, 79.2 years (28902 days), was obtained from a series of 77 estimates of the surface magnetic field $B_S$, obtained
over an interval of 10583 days \cite{2022MNRAS.514.3485G}.
Therefore, in what follows we will
use precisely this period estimate. The main parameters
of the star are taken from the catalog of fundamental parameters of Ap stars
\cite{2019AstBu..74...66G}: 
effective temperature $T_{eff} = 7520$ K,
surface gravity $\log(g) = 4.01$ cm s$^{-2}$, stellar radius in solar radii $R/R_{\odot} = 2.09$, stellar mass in solar masses $M/M_{\odot} = 2.04$, and
age $\log(t) = 8.98$ yr.
From estimates of
$B_S$ for 33 Lib with a period of 28902 days \cite{2022MNRAS.514.3485G},
an average magnetic phase curve and its parameters were obtained by the method described in \cite{2005A&A...430.1143B},
and are shown in Fig.~\ref{fig:MFK}.

The resulting MPC
is close to simple harmonic dependences, which makes it possible
to assume a dipolar structure of the global magnetic field
and to determine its parameters.
It should
be noted that
$B_S$ for this star varies only slightly with the
rotation period, and therefore it makes sense
to obtain an average estimate of the observed $B_S$ values from
data from \cite{1997A&AS..123..353M,2017A&A...601A..14M,
    2022MNRAS.514.3485G}.
The mean value of $B_S$ from 77 measurements is $4693 \pm 27$ G,
which is very close to the estimate of 4676 G given in the catalog \cite{2019AstBu..74...66G}.
Therefore, knowing from observations that
the value of $B_S$ varies little, for estimating the magnetic-field strength at the pole, $B_p$,
we can use the expression from \cite{1950ApJ...112..222S}
$B_p = BS/0.63$.
Using this expression, we obtain $B_p = 7450$ G, understanding that this is a rather approximate value.
The semi-amplitude of the MPC obtained from the
$B_S$ estimates is only 42 G, which may lead
to an uncertainty in the estimate of $B_p$ of $\pm67$~G.

Data from the TESS, ESPRESSO, and MOST projects continue to refine the role of magnetic braking in the dynamics of such objects \cite{Holdsworth2018}.

\subsection{Analytical Model}
\begin{figure}
    \centering
    \includegraphics[width=1.0\textwidth]{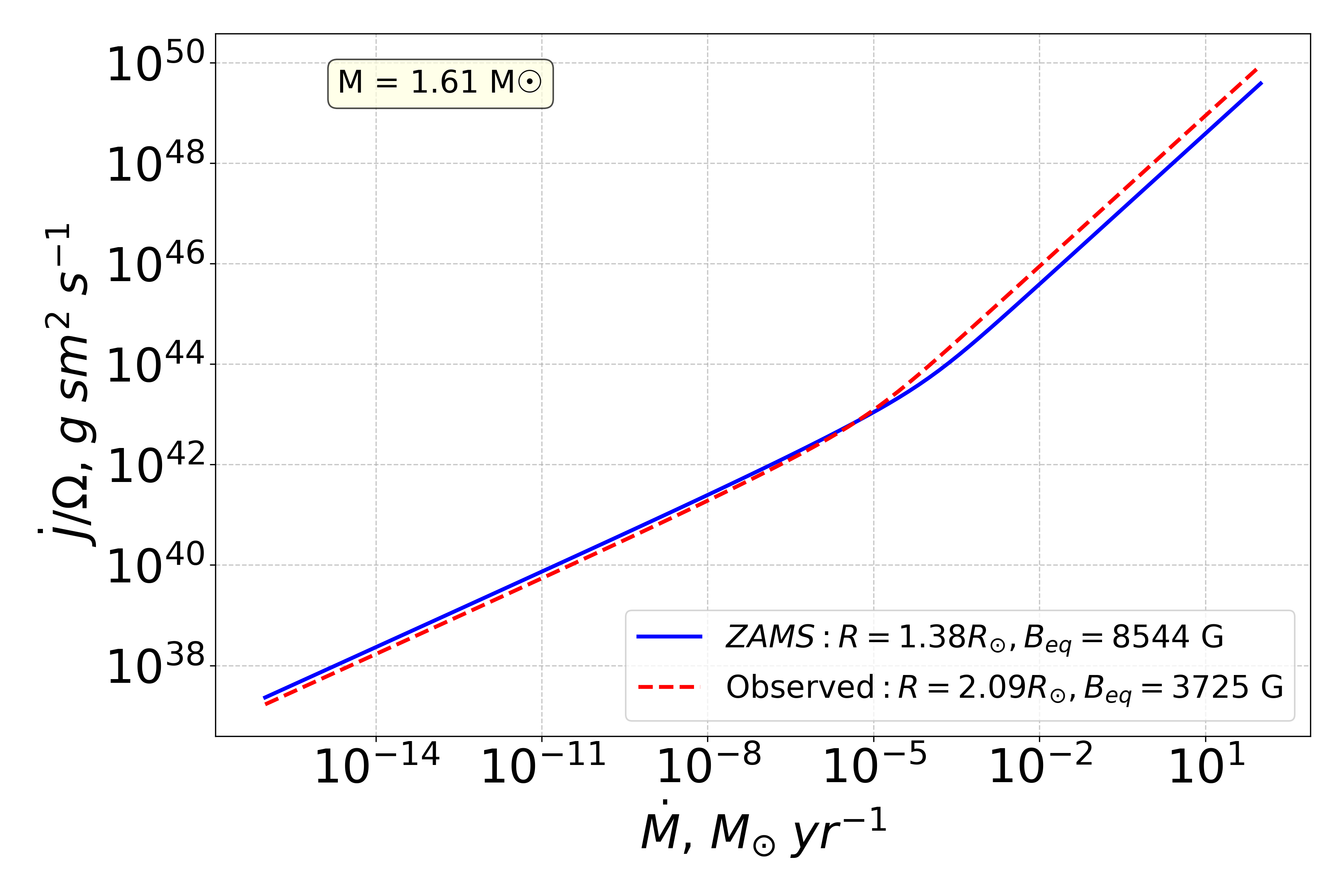}
    \caption{
    Parameterization of the analytical dependence of the angular-momentum loss rate $\dot{J}(\dot{M}_*)$ on the mass-loss rate $\dot{M}_*$, calculated from the Weber--Davis formula (Eq.~\ref{eq:WD}). Two fixed sets of stellar parameters obtained from \texttt{MESA} modeling are used for the analysis: for the initial stage on the main sequence (blue curve) and for the current observed state (red curve). The abscissa shows the mass-loss rate $\dot{M}_*$, and the ordinate shows the angular-momentum loss rate $\dot{J}$. Note: only $\dot{M}_*$ is varied in the calculation; the remaining parameters (angular velocity $\Omega$, Alfv\'en radius $R_A$) are fixed in accordance with the selected evolutionary state ($L=~11.77L_{\odot}$, $T=7400$ K).}
    \label{fig:jdot}
\end{figure}

Two-dimensional magnetohydrodynamic (MHD) calculations performed in \cite{uddoula2009} and the 3D MHD modeling confirming them \cite{2023MNRAS.520.3947U} show that for stars with a strong dipolar magnetic field the Alfv\'en radius can be calculated as
\begin{equation}\label{eq1}
    R_A \approx R_{s}(0.29+(\eta_* + 0.25))^{1/4},
\end{equation}
where $R_{s}$ is the stellar radius and $\eta_*$ is the wind magnetic-confinement parameter. This parameter is defined as the ratio of the energy flux of the electromagnetic field to that of the stellar wind near the stellar surface \cite{uddoula2009}:
\begin{equation}
    \eta_* \equiv \frac{B_{eq}^2 R_s^2}{\dot{M} v_{\infty}},
    \label{eq:eta}
\end{equation}
here 
$\dot{M}$ is the mass-loss rate and $v_{\infty}$ is the terminal velocity of the outflowing wind. One also defines the angular-momentum loss rate for the Weber--Davis dipole (dipole WD) \cite{weber1967, uddoula2009}
\begin{equation}
    \dot{J}_{dWD} = \frac{2}{3} \dot{M} \Omega R_A^2,
    \label{eq:WD}
\end{equation}
an illustrative curve is shown in Fig. \ref{fig:jdot}. For a main-sequence star with a strong magnetic field of about 18 kG, we see a break at $\dot{M}\approx 10^{-4} M_\odot$~yr$^{-1}$. Above this value, the effective Alfv\'en radius becomes constant and is of the order of the stellar radius, and the braking is proportional to the mass-loss rate, $\dot{J}\propto\dot{M}$. Below the critical value, the decrease in the mass-loss rate is compensated by the growth of the Alfv\'en radius, and the braking rate is proportional to $\dot{J}\propto B_{eq} \dot{M}^{1/2}$, which allows the star to be braked even at relatively small mass-loss rates.

\subsection{Numerical Model}

The \texttt{MESA} software package was used for numerical modeling of stellar evolution \cite{Paxton2011, Paxton2013, Paxton2015, Paxton2018, Paxton2019}.
There are no reliable observational data on the mass-loss rates of low-mass stars with radiative envelopes \cite{2014A&A...564A..70K}  
, so we applied the wind model from \cite{Nieuwenhuijzen1990}


\begin{equation}
        \dot{M} = 9.6 \times 10^{-15} \xi \left( \frac{L}{L_{\odot}} \right)^{1.42} \left( \frac{M}{M_{\odot}} \right)^{0.16}\left( \frac{R}{R_{\odot}} \right)^{0.81} \;M_{\odot}~\text{yr$^{-1}$},
\end{equation}
where $L$, $M$, and $R$ are the luminosity, mass, and radius of the star, respectively, and $\xi$ is a dimensionless parameter.
The use of this mass-loss model is determined by the effective temperature ($T_{eff}<10^4$ K) of the star HD137949 under study and by the fact that it is considered applicable to most main-sequence stars on the Hertzsprung--Russell diagram; however, as it turned out, this formula most likely strongly overestimates the mass-loss rate for moderate-mass stars with radiative envelopes ($1.25 M_\odot < M < 2 M_\odot$), i.e., it turned out that $\xi\ll1$; see the Results section for details.

\begin{figure}[h]
    \centering
    \includegraphics[width=1.0\textwidth]{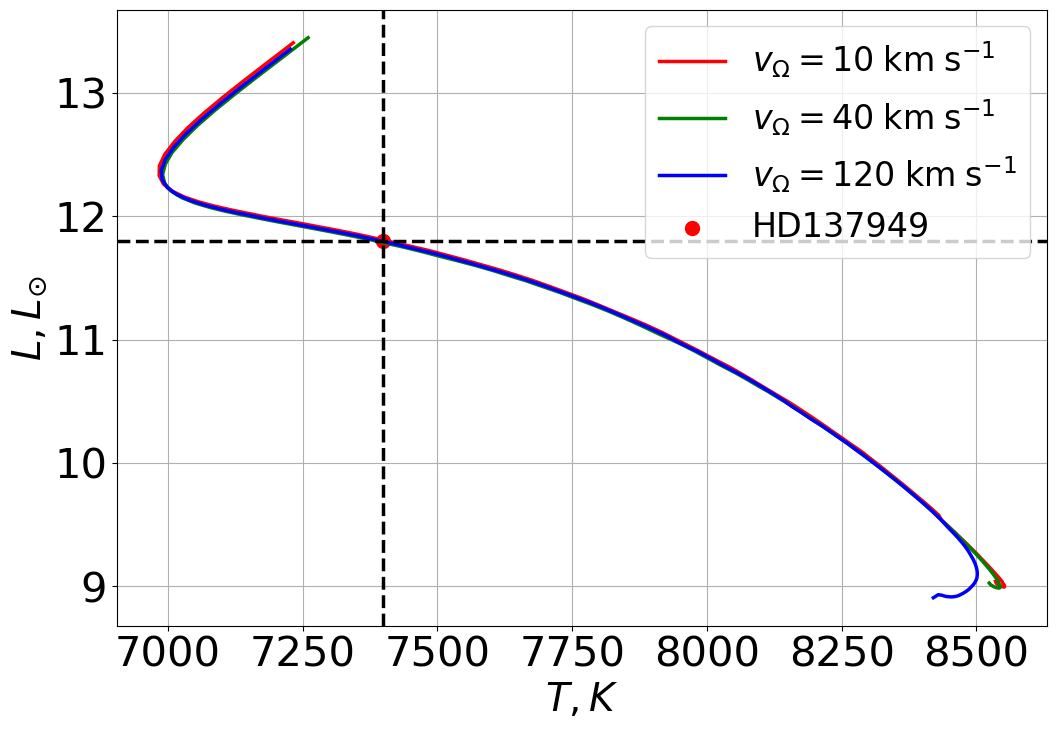}
    \caption{Evolutionary track of the star HD137949 for three different initial rotational velocities $v_{\Omega} = \{10, 40, 120\}$ km/s. The red point indicates the observed parameters of HD137949.}
    \label{fig:tracks}
\end{figure}

For the adopted parameters of the star under study, $L=~11.77L_{\odot}$, $T=7400$ K, $R=2.09R_{\odot}$ \citep{2013A&A...551A..14S}, we found a stellar mass of $M=1.606M_{\odot}$, which allowed the track on the HR diagram to pass through the observed point (Fig. \ref{fig:tracks}). Using the derived parameters and adopting $\xi=1$, we estimate the mass-loss rate as
\begin{equation}{\label{eq8}}
    \dot{M}_{0} = 6.24\times10^{-13} \quad {M}_{\odot}~\mbox{yr}^{-1}.
\end{equation}
In this work the notation $a_b \equiv a/10^b$ will often be used; all quantities are given in cgs units unless stated otherwise.

For a known mass-loss rate, one can write an analytical estimate of the evolution of the angular velocity. For this purpose, from a differential equation of the form
\begin{equation}
    \dot{\Omega}(t) = \chi_0 \Omega,
    \label{eq:Omega_dot}
\end{equation}
where
\begin{equation}
    \chi_0=\frac{2\dot{M}R_s^2}{3I_s} \left(0.29+\left(\frac{\mu^2}{R_s^4 v_w \dot{M}} +0.25\right)^{1/4}\right)^2,
    \label{eq:chi_zero}
\end{equation}
where $I_s$ is the total moment of inertia of the star, $I_s=\kappa M R_s^2$, $\kappa$ is a dimensionless moment-of-inertia coefficient that depends on the mass distribution inside the star, $\mu$ is the dipole magnetic moment, and $v_w$ is the stellar-wind velocity.
This differential equation is easily solved and gives an exponential solution of the form
\begin{equation}
    \Omega(t) = \Omega_0 e^{-\chi_0 t},
    \label{eq:omegae}
\end{equation}
where $\Omega_0$ is the initial angular velocity of the star. As follows from Eq.~(\ref{eq:Omega_dot}), the coefficient $\chi_0$ is the inverse of the braking time, $\chi_0=t_{sd}^{-1}$.
We also introduce the angular velocity corresponding to the currently observed 80-year period of the star, which we shall regard as the minimum rotation rate for subsequent calculations,
\begin{equation}
    \Omega_{p} = 2.5\times10^{-9} \qquad \text{rad}\cdot\text{s}^{-1}.
\end{equation}

The total moment of inertia for the object under study is
\begin{equation}{\label{eq11}}
    I_s = \kappa M R_s^2=10^{54} \kappa_{-1.5} M_{33.5} R_{11}^2 \qquad \text{g}\cdot \text{cm}^2,
\end{equation}
where $\kappa_{-1.5} = \kappa/ 10^{-1.5}$.
Let us estimate the braking time using Eq.~(\ref{eq:chi_zero}) in the approximation $\dot{M}~\ll~10^{-4}\;M_\odot/$~yr, as
\begin{equation}{\label{eq:tsd}}
    t_{sd} = \frac{3}{2} \frac{\kappa M}{B_{eq}R_s}\sqrt{\frac{v_w}{\dot{M}}}=1.5 \times 10^{15}\frac{\kappa_{-1.5} }{B_4 \dot{M}_{-12}^{1/2}} \frac{M_{33.5} v_8^{1/2}}{R_{11}}\qquad \text{s},
\end{equation}
where $B_{4}$ is the magnetic field in units of 10 kG, $\dot{M}_{-12}$ is the normalized mass-loss rate in units of $10^{12}$ g/s, and $v_{8} = v_w /10^8$; the wind velocity was taken to be $10^8$ cm/s.
Using the present-day parameters of the star under study and $\dot{M}~=~10^{12}$~g/s, the characteristic braking time is $t_{sd} \simeq 6\times 10^7$ yr.

\begin{figure}[h]
    \centering
    \includegraphics[width=1.0\textwidth]{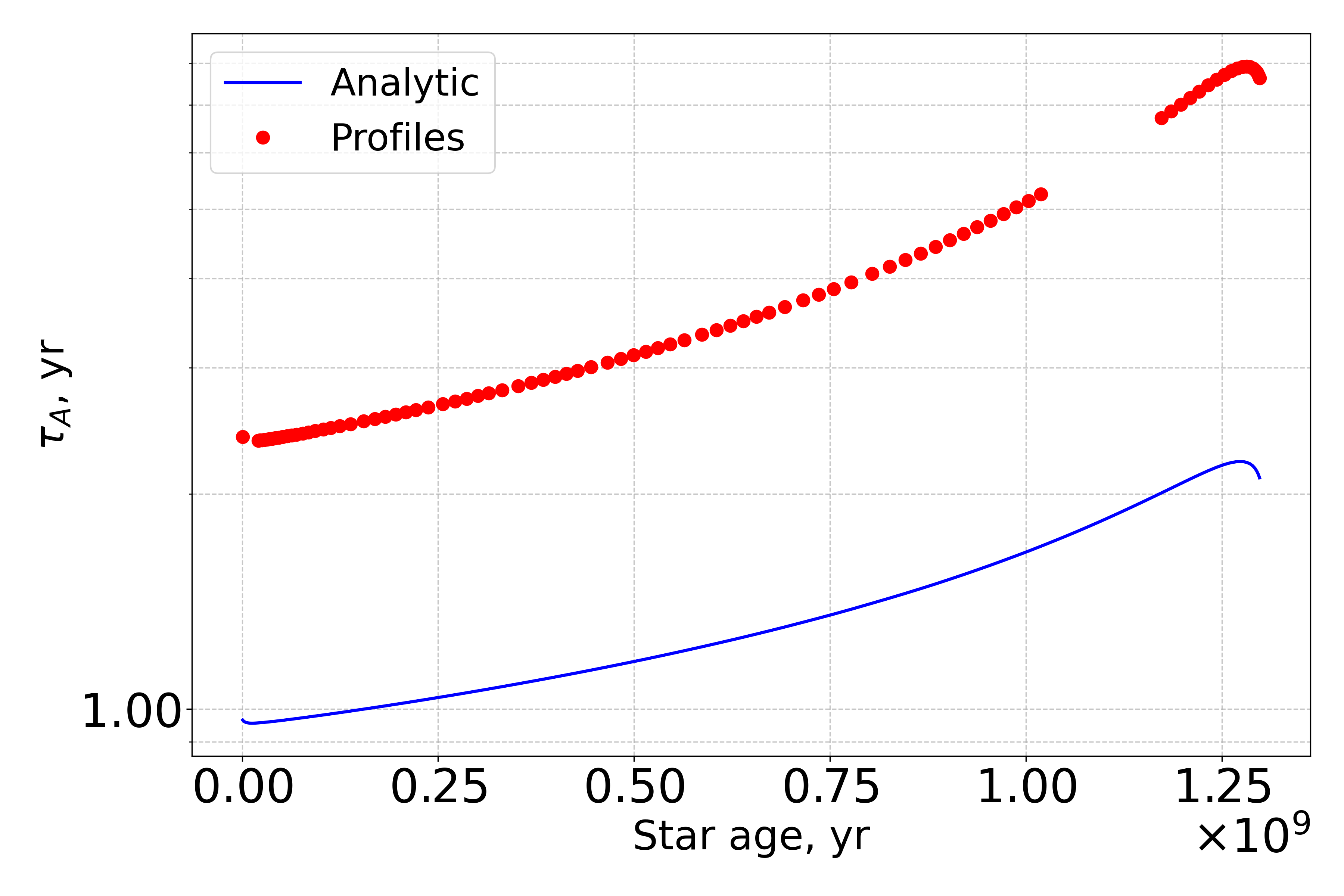}
    \caption{Variation of the Alfv\'en time over stellar evolution. The red points show the Alfv\'en time calculated from the integral density profiles of the star. The blue line shows the analytical approximation. It is seen that during stellar evolution the Alfv\'en time does not exceed $\tau_A \le 8$ yr.}
    \label{fig:Alfven_time}
\end{figure}

To justify the use of the solid-body model, the Alfv\'en time $\tau_A$ was estimated as
\begin{equation}
\tau_A(t) = \frac{R_s(t)\sqrt{4\pi\bar{\rho}_*(t)}}{B_p(t)}= 3.5\times 10^7\frac{M_{33.6}^{1/2}}{B_4 R_{11}^{1/2}} \qquad \text{s} ,
    \label{eq:alfven_time}
\end{equation}
where $R_s(t)$ is the stellar radius and $\bar{\rho}_*$ is the mean stellar density. Its evolution is shown in Fig.~\ref{fig:Alfven_time} (solid line), calculated using the \texttt{MESA} package with nonstationary parameters.

A comparison of the Alfv\'en times computed from the stellar evolutionary profiles, for which the radial density distribution is known, and from the analytical approximation under the assumption of a star with uniform density is presented in Fig.~\ref{fig:Alfven_time}. Here
\begin{equation}
    \tau_A = \int_0^{R_s(t)} \frac{\sqrt{4 \pi\rho(r)}}{B(r(t))}dr,
     \label{eq:Alfven_time}
\end{equation}
where $\rho(r)$ is the density profile at specific moments in time, and $B(r(t))$ is the corresponding magnetic-field value at those moments, taking magnetic-flux conservation into account:
\begin{equation}
    B(t)=B_c\left(\frac{r(t)}{R_s(t)}\right)^2,
    \label{eq:magn_flux}
\end{equation}
where $B_c$ is the currently observed magnetic-field value, and $r$ and $R_s$ are the current radial coordinate and the radius of the stellar photosphere, respectively. For conserved magnetic flux, with the observed values $B_{p,c} = 7.4 $ kG and $R_{s,c} = 2.09R_{\odot}$, computed for a mass $M_s=1.606M_{\odot}$, the initial stellar radius was $R_{s,0}=1.38R_{\odot}$, which corresponds to an initial magnetic field of $B_{p,0} = 17.1$ kG. For a dipole, the surface magnetic-field strength at the magnetic equator is $B_{eq} = B_p/2$.

In Fig.~\ref{fig:Alfven_time}, the dependence of the Alfv\'en time $\tau_A$ calculated from Eq.~(\ref{eq:alfven_time}) is shown by the blue line, while the value computed from the evolutionary model, as shown in Eq.~(\ref{eq:Alfven_time}), is shown by the red points.
The difference between the times calculated from the mean density and from the evolutionary density profiles is approximately a factor of two. This is due to the strong concentration of mass toward the center in the numerical model, which leads to a strong decrease in the propagation speed of Alfv\'en (torsional) waves.
In any case, the Alfv\'en time does not exceed a few years. Thus, for the star 33~Lib the Alfv\'en time is always much shorter than the magnetic braking time $t_{sd}$ and the current stellar rotation period $P=80$ years.

\subsection{Numerical Simulation}

In the numerical calculation with \texttt{MESA}, we apply the solid-body approximation.
This makes it possible to neglect the influence of magnetic fields on the large-scale structure and evolution of the star, since the slow development of magnetic effects does not have time to change the system significantly. By simplifying the model through the exclusion of complex magnetohydrodynamics, the solid-body approximation significantly reduces the computational cost while preserving accuracy in scenarios dominated by gravity, thermal effects, or stellar wind.\footnote{ In the configuration files of the \texttt{MESA} code, \texttt{inlist\_braking}, solid-body rotation used \texttt{set\_uniform\_am\_nu\_non\_rot = .true.} in the \texttt{\&controls} section.}

\begin{figure*}[htbp]
    \centering
    \includegraphics[width=1.0\textwidth]{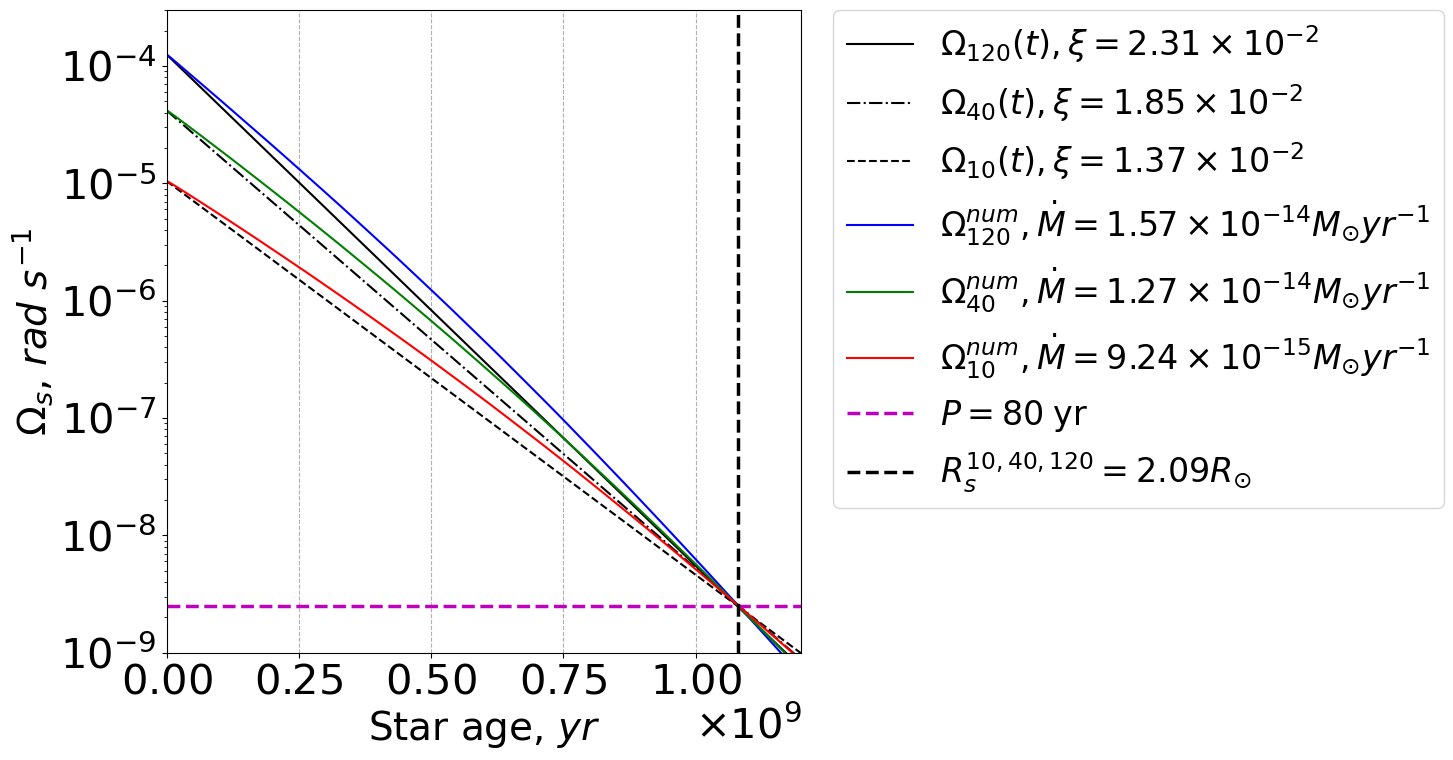}
    \caption{Change in the stellar angular velocity over the course of evolution for a variable dipolar magnetic field and different initial values. The initial values of the equatorial linear rotation velocity are $v_\Omega= \{10, 40, 120\}$ $\text{km} \;\text{s}^{-1}$.
   The upper rotation-velocity limit is set as $v_{\Omega,\text{max}} = \frac{1}{3}v_{\text{Kep}}(R_0)$, where $R_0~\simeq~10^{11}$~cm is the initial stellar radius. This restriction, which is more conservative than the estimates of the critical rotation velocity in \cite{2010RMxAC..38..113M}, is introduced in order to suppress significant meridional mixing. Here the mass-loss rate $\dot{M}$ is indicated for the stellar rotation period $P=80$ yr.}
    \label{fig:Angular_velocity}
\end{figure*}

In the numerical calculation, the free parameter is the mass-loss efficiency coefficient $\xi$; by varying it one can obtain the required rate of spin-down to the observed physical parameters of the star, radius $R=2.09R_{\odot}$ and rotation period $P=80$ yr.
Figure \ref{fig:Angular_velocity} shows the angular velocity of the star as a function of age. In the model under consideration, a dipolar magnetic-field configuration with constant magnetic flux throughout the evolution is adopted. Assuming magnetic-flux conservation (Eq.~\ref{eq:magn_flux}), we considered three models with different values of the initial equatorial linear rotation velocity of the star, $v_{\Omega,0} = \{10,40,120\}$ km $\text{c}^{-1}$.
For the analytical curves $\Omega_{i}(t)$ in Fig.~\ref{fig:Angular_velocity}, shown by black lines, the free parameter was the mass-loss rate $\dot{M}$ or the wind parameter $\xi$.

\begin{figure}[h]
     \centering
     \includegraphics[width=1.0\textwidth]{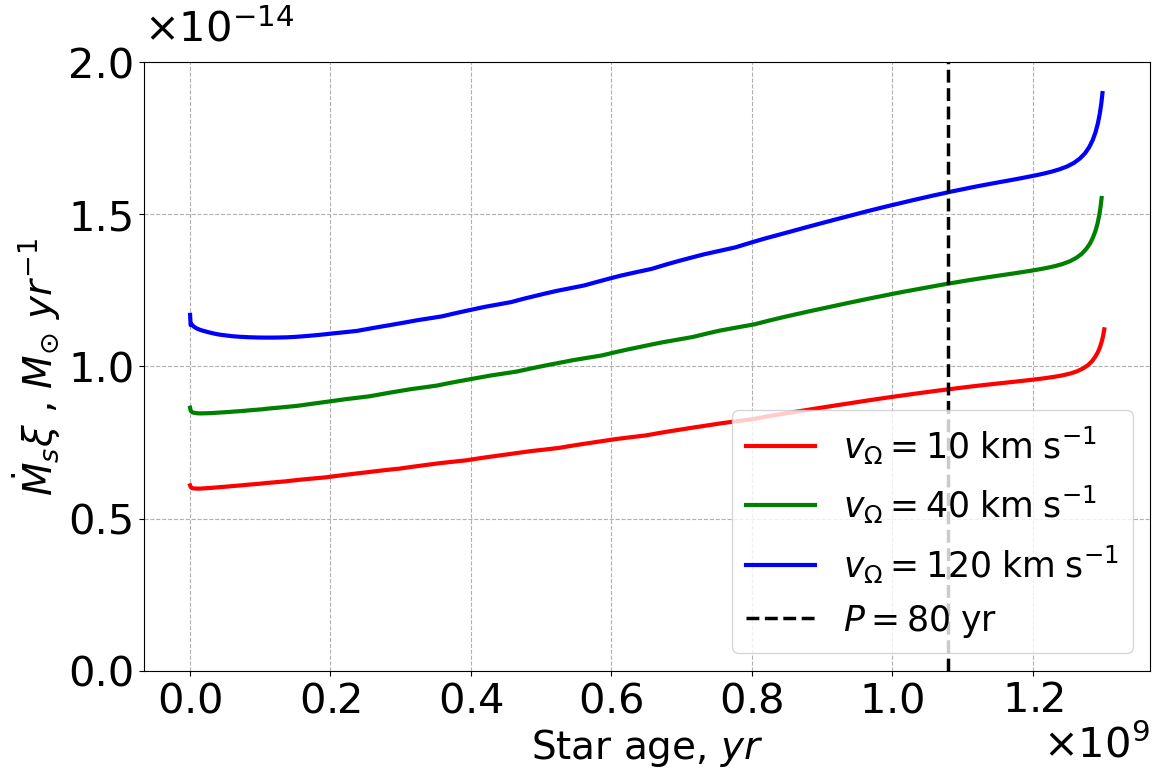}
     \caption{Variation of the mass-loss rate of the star HD137949 for initial rotational velocities $v_\Omega= \{10, 40, 120\}$ km $\text{s}^{-1}$. The ordinate shows the mass-loss rate $\dot{M}_s$ in solar masses per year, and the abscissa shows stellar age over the course of evolution. The change in the mass-loss rate is shown for different values of the wind efficiency $\xi$ (see Fig.~\ref{fig:Angular_velocity}) that provide stellar braking down to a rotation period $P=80$ yr and a stellar radius $R_s=2.09R_{\odot}$.}
     \label{fig:varmdot}
 \end{figure}

In the calculations, conservation of magnetic flux over the entire lifetime of the star was taken into account, which led to changes in the magnetic-field strength during stellar evolution (see Eq.~\ref{eq:magn_flux}). Thus, during the evolutionary calculations the radius, magnetic field, luminosity, and consequently the mass-loss rate changed self-consistently (see Fig.~\ref{fig:varmdot}).

\begin{figure}
    \centering
    \includegraphics[width=1.0\textwidth]{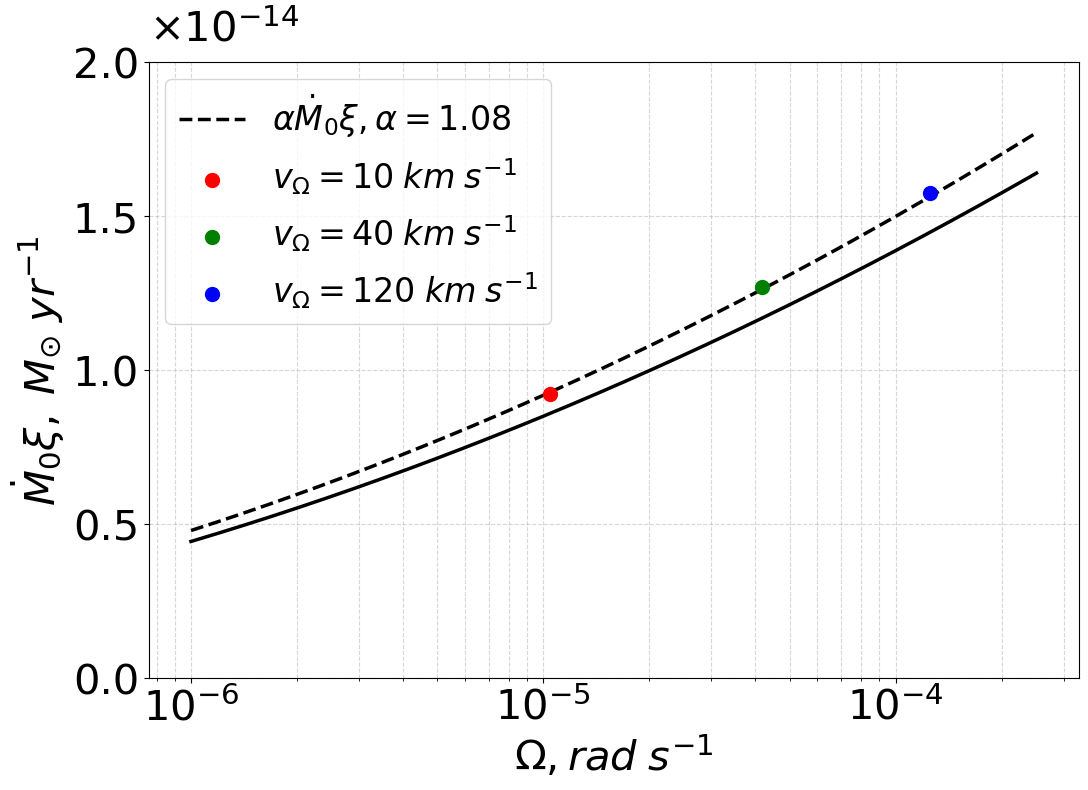}
    \caption{Dependence of the mass-loss rate on the value of the angular velocity for the parameters of the observed star HD137949. The solid line shows the theoretical dependence $\dot{M}_0 \xi$. The dashed curve shows the theoretical dependence with a dimensionless coefficient $\alpha =1.08$ to describe the results of the numerical modeling, which are indicated by points. }
    \label{fig:Mxi_omega}
\end{figure}

\begin{figure}[h]
    \centering
    \includegraphics[width=1.0\linewidth]{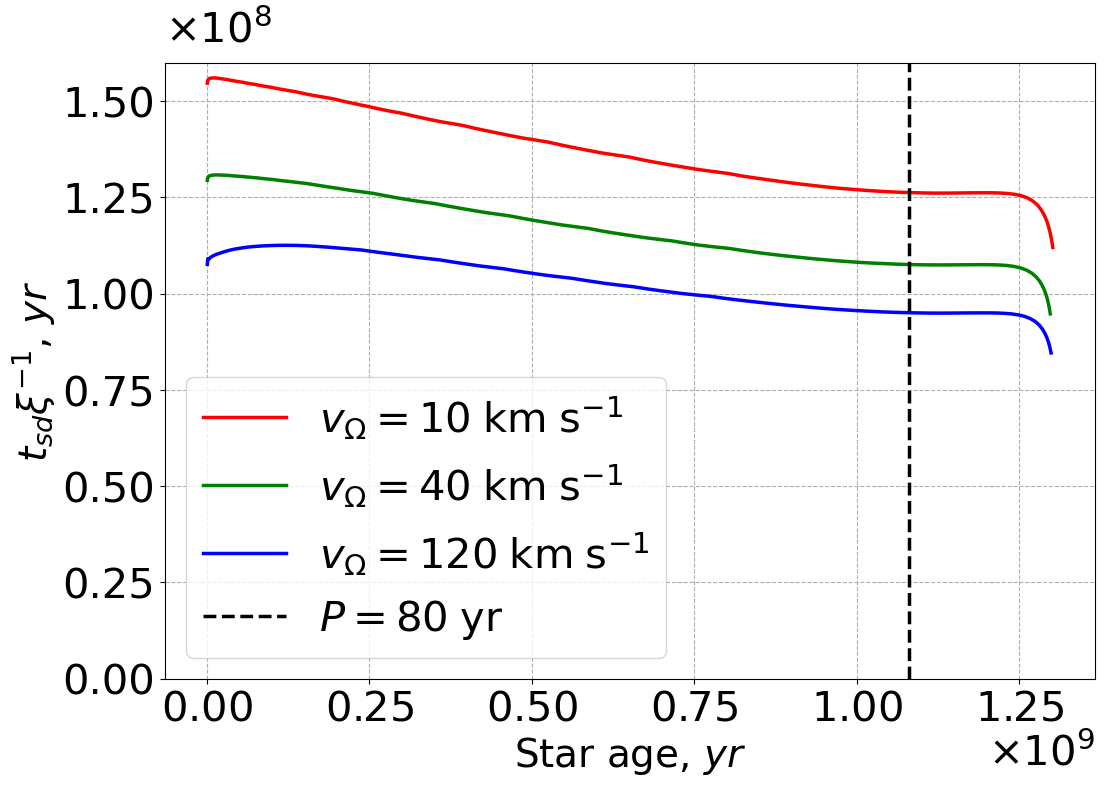}
    \caption{Magnetic-wind braking time for the star HD137949 for different values of the initial rotational velocity. The ordinate shows the braking time $t_{sd}$ in years, and the abscissa shows stellar age over the course of evolution. The dashed line indicates the age at which the rotation period reaches $P=80$ yr.}
    \label{fig:spindown_time}
\end{figure}

In Fig.~\ref{fig:Mxi_omega}, the solid line shows the theoretical dependence of the mass-loss rate on the initial angular velocity of the star that is required to brake the stellar rotation to the observed values; it was obtained from expression (\ref{eq:omegae}). The mass-loss rates obtained from the numerical modeling are shown by points for different initial rotational velocities. As can be seen from the plot, the numerical modeling gives only a minor deviation of the wind estimate compared with the theoretical one (the difference is a factor of $1.08$). In the analytical model, the evolution of the radius over the star's main-sequence lifetime is neglected, which affects the strength of the surface magnetic field and the stellar moment of inertia.
Fig.~\ref{fig:spindown_time} shows the magnetic wind braking time $t_{sd}$ for HD 137949 at different initial rotation velocities, depending on the star's age and its evolution stage. The vertical dashed line indicates the age at which rotation period is $P = 80$ years.

\section{Results and Discussion}

The magnetic-field topology of the star 33 Lib was assumed to be dipolar, which is consistent with the MPC data, and the field was assumed to be aligned with the stellar rotation. This assumption is supported by 3D MHD calculations \cite{2023MNRAS.520.3947U}, which showed that an oblique rotator, to within an accuracy of about 30\%, has the same braking rate as an aligned rotator.
A more complex magnetic-field topology was not considered, because a dipole field decreases with distance most slowly and, accordingly, provides the largest Alfv\'en radius and the most efficient braking. If the magnetic field contains higher-order multipoles, then the observed MPC will be described by trigonometric polynomials, while the observed segment of the curve covers only its rising part (see Fig.~\ref{fig:MFK}). In other words, for a fixed derivative, the periodic timescale of the polynomial will be longer than in the case of a simple sinusoid, which arises for a dipole field. Thus, MPC observations would be explained by rotation periods many times longer than 80 years.

In the case of braking by a magnetized stellar wind, expression (\ref{eq:omegae}) obtained in this work shows that the angular velocity decays exponentially over the course of stellar evolution, and for stars with strong magnetic fields the star can practically stop or slow down substantially. It turned out that even a low-intensity stellar wind ($\dot{M}\sim10^{-14}M_\odot$~yr$^{-1}$) is capable of efficiently braking ssrAp stars down to the observed rotation rate of a star such as 33 Lib.

We propose introducing a dimensionless parameter that is, in essence, the ratio of the main-sequence lifetime to the magnetic-wind braking time:
\begin{equation}
    \Xi = \frac{\dot{M}_{-12}^{1/2} B_{eq,3}}{M_{33.5}^{2.4}} \frac{1}{\kappa_{-1.5} \sqrt{v_{w,8}}}
\end{equation}
for values larger than unity, according to a simple estimate made using Eq.~(\ref{eq:tsd}), the braking time of an A-type star is of the order of its lifetime. In the future, we plan to carry out statistical studies of magnetic stars.

The present study demonstrates that magnetic braking caused by the interaction of a strong magnetic field and a stellar wind can play a key role in slowing the rotation of ssrAp stars. Modeling of the evolution of magnetic fields with a variable magnetic configuration and stellar wind confirms the simple theoretical model. A magnetized wind can efficiently brake rotation, leading to the observed slow rotation of A-type stars.


The procedure presented in this paper, given a known magnetic field, may prove to be a unique method for measuring the mass-loss rate of A-type stars, which may be difficult to determine by other means.


\clearpage
\section*{Acknowledgments}
The authors express their deep gratitude to Yu.A.~Fadeev, T.A.~Ryabchikova, and L.I.~Mashonkina for productive discussions of the materials of this work.
The work of A.G.N. and M.V.B. was supported by BASIS Foundation grant \#24-1-2-25-1 and was carried out within the framework of the state assignment of the Institute of Astronomy of the Russian Academy of Sciences.

\section*{CONFLICT OF INTEREST}
The authors declare that they have no conflict of interest.

\clearpage
\bibliographystyle{maik}
\bibliography{article2}

\end{document}